\documentstyle[sprocl]{article}
\def\Journal#1#2#3#4{{#1} {\bf #2}, #3 (#4)}
\def\be{\begin{equation}}
\def\ee{\end{equation}}
\def\bea{\begin{eqnarray}}
\def\eea{\end{eqnarray}}
\begin{document}
\title{THE DISTRIBUTION OF DARK MATTER IN THE MILKY WAY GALAXY}
\author{ GERARD GILMORE}
\address{Institute of Astronomy, Madingley Rd, \\ Cambridge CB3 0HA, UK}

\maketitle\abstracts{A wealth of recent observational studies shows
the dark matter in the Milky Way to have the following fundamental
properties:
\item{1)} there is no detectable  dark matter associated
with the Galactic disk -- the dark matter is distributed in a purely halo
distribution with local volume density near the Earth $\approx 
0.3$GeV/cc $\equiv 0.01M_\odot$pc$^{-3}$;
\item{2)} The stellar mass function is both universal and convergent at
low masses -- there is no significant very low luminosity baryonic
dark matter associated with stellar light. Fundamentally, dark mass does
{\em NOT} follow light.
\item{3)} The smallest scale length on which dark matter is clustered
is in the Milky Way's satellite dwarf Spheroidal galaxies, where dark
matter has a characteristic length scale $\sim 1kpc \equiv 10^{20}m$
and mass density $\sim 0.05M_\odot$pc$^{-3} \equiv  1.5$GeV/cc.
}

\section{Dark Matter In The Galactic Disk: An Update}

\subsection{The Kuijken-Gilmore Standard determination of the local
dark matter density}

The standard determination of the volume density of matter near the
Sun is that due to Kuijken and Gilmore (Kuijken \& Gilmore
1989a,1989b,1989c,1991). Kuijken and Gilmore developed a new maximum
likelihood technique to analyse the joint kinematic and spatial
distribution functions, under the constraint of physical
realisability, which they applied to new photometric and kinematic
data. Their analysis utilised their kinematic and photometric survey
of K dwarf stars towards the South Galactic Pole, perpendicular to the
Galactic Plane. The essential aspects of their method derive
from the fact that determination of the mass distribution of the
Galactic disk from stellar kinematic tracers essentially requires
comparison of the velocity distribution function at some height from
the disk mid-Plane with the spatial density of the same tracer
population above that height.

For this reason, purely local determinations of the volume mass density are
inherently inaccurate, as they must compare both local all-sky and
distant pencil-beam data. The Kuijken and Gilmore analysis utilised
velocities for stars up to $\sim 1$kpc from the Galactic Plane,
thereby minimising dependence on poorly determined local
quantities. This provided a robust determination of the total,
integral, {\em surface mass density} of all gravitating matter within
1.1kpc of the Galactic Plane, which is $71\pm6 M_\odot pc^{-2}$.  By
combining this result with large scale constraints on the mass
distributions in the disk, bulge and halo, from the Galactic rotation
curve, Kuijken and Gilmore determined the column integral mass density
of the Galactic disk to be $48\pm9 M_\odot pc^{-2}$.

The question imediately arises as to what fraction, if any, of this
total disk mass might be (cold) dark matter distributed with a
disk-like density distribution. Recall that the 1-D velocity
dispersion associated with the Galactic disk density distribution is
$\sim 30km/s$, so any disk-like dark matter must indeed be cold dark
matter. There is no reason to expect any mass with such a cold
distribution to be related to the mass which dominates the large scale
galactic potential of course. Direct observational measurement of the
identified baryonic disk mass is not easy. Kuijken and Gilmore (1989b)
describe the contributions and methods, while Gilmore, Wyse  and
Kuijken (1989) summarise the situation in the wider, Galactic,
astrophysical context.  Basically, one must use the local observed
stellar mass function and kinematic distribution function, integrated
through the gravitational potential determined by Kuijken and Gilmore,
to determine the stellar contribution to the column integral. The
contribution from atomic and molecular gas is then added, to provide
the total. Proceeding in this way, Kuijken and Gilmore calculated the
identified baryonic mass surface density of the Galactic disk to be
$48\pm8 M_\odot pc^{-2}$.

Thus, there is no evidence for any significant unexplained mass
associated with the Galactic disk. Rather, the mass required to
support the Galactic rotation curve, in
addition to the identified disk and bulge mass, must be distributed in an
extended halo.  It is this mass whose local {\em volume} density is 
 $\approx~0.3$GeV/cc~$\equiv~0.01M_\odot$pc$^{-3}$, and which has a
velocity distribution approporiate to an extended halo distribution,
which is the target of the particle dark matter searches described
elsewhere in this volume.

\subsection{Developments subsequent to the Kuijken/Gilmore
experiment}

There are four aspects of the determination of the Galactic
Gravitational Force Law ${\cal K}_z(z)$ which can be tested using more
recent data. The adopted distribution of chemical elements in the
tracer stars; correctness of the kinematics; correctness of the photometry, and
calculation of the identified baryonic disk mass.

1) Potentially the most important is the stellar chemical abundance
distribution function far from the Galactic Plane. Distances for the
tracer stars, and hence the distance scale of the experiment, are
determined essentially from the Hertsprung-Russell diagram, which
relates stellar colours to intrinsic luminosity. Thus, determination
of a stellar colour and the inverse square law are the primary scale
factors. Stellar colours are however also systematically dependent on
the abundances of the chemical elements, a distribution which was very
poorly known, for stars far from the Galactic plane, at the time of
the Kuijken/Gilmore experiment. Should the abundance distribution
function adopted be systematically in error, then a systematic error
would ensue in the derived force law, and generating mass ditributions.

Determination of stellar abundances far from the Plane is a
technically challenging  and
time-consuming task in its own right. An extensive project to improve
its determination has recently been completed by Wyse and Gilmore
(1995), utilising techniques and results from Jones, Gilmore and Wyse
(1996), Jones, Wyse and Gilmore (1995), and Gilmore, Wyse and Jones (1995).
More recently, further support for this result has
been provided from analyses by B. Nordstrom and J. Andersen
(pri. commn) based on detailed studies of very large samples of bright
stars, and by Reid, Gizis, etal (in preparation), based on smaller
samples of very faint low-luminosity stars. 
The abundance distribution function derived in these projects, from careful
analysis of spectra and kinematics of large samples of stars, are in 
surprisingly (to this author) good agreement with that adopted by Kuijken and
Gilmore (1989b). 

2) Measurement of radial velocities using the multi-plex
optical fibre technique utilised by Kuijken and Gilmore was at the
time a new technology. Since that time, both the methodology  (eg
Ibata and Gilmore 1995) and the detailed results in this line of sight,
have been confirmed by several authors (eg Gould, Bahcall and Flynn
1996 and refs therein). The kinematics seem reliable.

3) The Kuijken and Gilmore sample of tracer stars, and the
corresponding spatial density profile of these stars, were derived
from photograpic photometry, inevitably requiring several large
systematic corrections for conversion to a calibrated scale. The
relevant photometric techniques are complex, and require considerable
corrections during their application. A description is provided by
Gilmore (1983). Recent technological developments now allow
appropriate surveys with sensitive, linear, photometric detectors,
obviating the need for complex photographic photometry. The first
independent test of the results of the photographic technique has been
provided by a massive survey, using contemporary linear CCD digital
detectors, by Caldwell and Schecter (1996). These authors show the
Kuijken/Gilmore photometry to be accurate to within 2percent in zero
point, and 3 percent in colour calibration, in the range of
interest. Thus the basic photometric data are of sufficient precision
for the conclusions to be unaffected.

4) Potentially the least reliable part of the determination of the
density of dark matter near the Sun is the calculation of the actual
baryonic mass density associated with the Galactic disk.  This
calculation involves a conversion from the observed {\em luminosity}
distribution of stars to their {\em mass} distribution: a conversion
which is poorly known, and difficult to check.  Pending an eventual
future agreement among the very many groups modelling the
luminosity-mass conversion, we note here that the most recent
available determinations all agree, within better than two sigma, that
the identified mass density is indistinguishable from the dynamical
mass density. Among the very many relevant papers supporting this
basic result are those by Kroupa, Tout and Gilmore (1990, 1991, 1993),
Gould, Bahcall and Flynn (1996), Mera, Chabrier and Baraffe (1996) and
Chabrier (this volume).

Thus, all the very many aspects of the evidence are consistent with
the statement that there is no statistically significant evidence for
the existence of {\em any } dark matter {\em associated with the Galactic
disk}. Inside the Milky Way Galaxy, there is no evidence that dark
matter can concentrate on scales shorter than $10^{20}$m, and a
considerable body of evidence that it is collected only on much longer
scales. The full implications of this result for the velocity
dispersion and equation of state of the dark matter remain to be explored.

\subsection{Future improvements}

Is a future improvement plausible? Yes indeed. The ESA HIPPARCOS
astrometric satellite, the results of which will become available in
early 1997, will provide both direct trigonometric calibration of
distance scales and very accurate kinematics. A very substantial
improvement in the precision of the relevant measurements is
certain. Preliminary results (Pham 1997) strongly indicate that there
is indeed no significant dark matter distributed like the luminous
mass in the Galactic disk.  The galactic dark matter is a property of
the Galactic halo.

\section {Does Halo Dark Matter Follow Light?}

An enduring interest in studies of dark matter is the extent of a
baryonic contribution, associated with a high mass-light ratio part of
the stellar mass function (Lynden-Bell \& Gilmore 1990). This might
include, among more exotic possibilities, very low mass M-dwarf or
brown dwarf stars, or very old, cool, extremely low-luminosity, white
dwarfs. From the work of Reid and Gilmore (1982), utilising the new
automated photographic measuring machines and Schmidt telescope
surveys, to the ultra-deep Hubble Space Telescope survey of the Hubble
Deep Field (Elson, Gilmore, \& Santiago 1996), the limits in apparent
luminosity of surveys for such low luminosity sources have been
increased by a factor approaching one million. With every increase in
experimental sensitivity, the possibility of a substantial population
of low-mass, low-luminosity stars - always marginally significant, at
the limits of sensitivity, at best - has receded. Recent work,
including the recent gravitational microlensing results (Rich, this
volume, Sutherland, this volume), together with the first detection of
a real brown dwarf (Kulkarni, this volume) now provide convincing
evidence that very low mass stars can indeed be found, that their
volume demsity is very low, and that they are irrelevant for galactic
gravitational potentials.

The contribution of low mass stars to the total mass budget is
determined through the stellar mass function, the relative number of
stars per unit mass which are ever formed. Recent experimental and
theoretical results tightly constrain both the functional form of, and
possible variations in, this function.

\subsection{ Stellar mass functions in many environments}

Recent results from the Hubble Space telescope have substantially revised
and improved understanding of the stellar mass function. Accurate
determination of cluster stellar {\em luminosity } functions are now
available from several authors, for in total 2 open clusters and 11
globular clusters. In addition, many authors have determined the field
star luminosity functiom, with excellent agreement between independent
analyses (cf, eg Elson, Gilmore \& Santiago 1996; Flynn, Gould \&
Bahcall 1996; Gould, Bahcall \& Flynn 1996; Mendez, Minnitti, deMarchi,
Baker \& Couch 1996; Reid, Yan, Majewski, Thompson \& Smail 1996;
Santiago, Elson \& Gilmore 1996).

Analysis of these very many observational result shows excellent
agreement, even though a very wide range in chemical element
abundances and physical environment is considered. To excellent
precision, all {\em luminosity } functions have a broad maximum at a
luminosity which corresponds to a {\em mass} of $0.25M_\odot$ (Kroupa,
Tout \& Gilmore 1993; Santiago, Elson \& Gilmore 1996; von Hippel,
Gilmore, Tanvir, Robinson \& Jones 1996). 

Conversion of these luminosity functions to mass functions at the
lowest masses remains uncertain. Recent results however suggest a
stellar mass function which is perhaps slowly rising at very low
masses, but with a sufficiently shallow slope that no significant mass
remains to be accounted for (Kroupa, Tout \& Gilmore 1993; von Hippel,
Gilmore, Tanvir, Robinson \& Jones 1996; Gould, Bahcall \& Flynn 1996;
see also Chabrier, this volume). Thus, the stellar mass function is probably
universal, and has no dynamically significant mass at very low
masses. Low mass stars are irrelevant for the dark matter which
dominates galactic halos, and the Universe on large scales.

We emphasise here that this is no more than the well-known and usually
ignored result derived from elementary kinematic analyses: the
distribution of dark matter derived from rotation curve analyses is
fundamentally different, in every case studied, from the distribution
of the (stellar, baryonic) light. Mass does {\em NOT} follow light in
galaxies: total mass is invariably much more extended than is the
baryonic mass which generates the stellar light. Thus, adjusting the
mass to light ratio of any identified contribution to the observed
luminosity profile cannot, in principle, ever explain dark
matter. Matter which is distributed in space inherently differently
from all detected sources of luminosity is required to explain the
kinematic observations.

We thus are left with two questions concerning the distribution of
dark matter in the galaxy: is there a form of baryonic
dark matter distributed differently than the luminous stellar mass,
and so forming the Galactic halo? Does dark matter cluster on any detectable
scale smaller than the whole galaxy, so that we may constrain its nature?

\section{Does Baryonic Matter Contribute to the Galactic Dark Halo?}

If the answer to this question were known there would be no need for
this meeting. Substantial constraints are however rapidly
emerging. The EROS microlensing experiemnt effectively excludes
compact objects with masses below the stellar hydrogen burning limit
from contributing to the Galactic halo (Rich, this meeting). At
significantly higher masses, normal stars would be readily visible,
and have long been excluded. Two possibilities have remained: low
mass, but nonetheless hydrogen-burning, stars, too faint to have been
detected in available surveys; and very old white dwarfs. 

\subsection{Low Mass Stars as Halo Dark Matter?}

Both these possibilities can be tested with the new, ultra-deep Hubble
Space Telescope surveys, especially that in the Hubble Deep Field (eg
Elson, Gilmore, \& Santiago 1996), and from the wider-area Medium Deep
Survey Key Project (Santiago, Elson \& Gilmore 1996). The several
relevant analyses of these, and similar HST data, are in excellent
agreement, and limit the possible contribution of low mass normal
stars to the Galactic halo mass to being at an insignificant level,
with at most a few percent (Santiago, Gilmore \& Elson 1996) of the
mass which generates the rotation curve being allowed in this
form. Low mass stars are no longer viable as a substantial part of the
Galactic dark halo mass. Something more exotic is implicated.

\subsection{Old White Dwarfs as Halo Dark Matter?}

White dwarf stars are the cooling remnants of the cores of
intermediate-mass stars after they have exhausted their internal
energy sources. Their cooling rate is a weak function of
mass, with faster cooling, and hence lower predicted luminosities, at
higher masses. While white dwarfs from the normal disk and halo stellar
populations  all have masses very near
0.6$M_\odot$, the mass function for a new, hypothetical, population
may be different. After suitably long times, white dwarf remnants can attain
luminosities even lower than those of the lowest-mass hydrogen-burning
stars, and so can be very hard to detect observationally. Could a
population of very old white dwarf stars descended form a parent
population of normal stars make up the Galactic halo? This is
extremely unlikely.

There are several relevant astrophysical constraints. Some constraints
derive from the fact that a huge number of parent stars would have
been necessary. There is no evidence from surveys for luminous objects
at high redshift -- quite the reverse -- that any such massive star
formation phenomenon was universal in the Universe at very high
redshifts. And recall that no explanation for the dark matter halo of
the Milky Way Galaxy which requires special circumstances is viable:
dark matter is universal, so that its explanation must be fundamental to
the nature of the Universe at large.

Some constraints derive from the very large production of chemical
elements heavier than hydrogen which must have accompanied the lives
of the parent stars of the putative white dwarfs: all available
evidence is that the earliest generations of stars known - those which
must have formed {\em after} the putative early population - have
extremely low chemical abundances. The oldest stars known in the Milky
Way, those which form teh stellar halo, cannot have had precursors.  Some
constraints derive directly from the deep star counts outlined above:
one simply does not see any plausible candidate very old white dwarfs
(Elson, Gilmore \& Santiago 1996).

An interesting new limit, which seems to
exclude the white dwarf hypothesis effectively, has been derived by
Fuchs \& Jahreiss (1997). They calculate the local, Solar
neighbourhood, number of high-velocity, low-luminosity, white dwarfs
which must exist if the old white-dwarf halo model is viable. Since
such stars are in the immediate Solar neighbourhood, they are
sufficiently apparently luminous to be visible, irrespective of age. 
Additionally,  since they form an extended halo distribution, they have
extreme halo kinematics, hence are readily recognisable from their
space motions. Comparing this required number with that observed,
Fuchs and Jahreiss can exclude old halo white dwarfs as a viable model
for the Galactic halo.

We now consider what positive information concerning the distribution
of dark matter can be derived from stellar kinematics in the Galactic
satellites.

\section{ Does Dark Matter Cluster on Short Length Scales?}

Dark matter can be detected in principle on any length scale for which
a suitable kinematic tracer is available. The shortest relevant length
scale is the few parsec size of globular clusters. Globular clusters
show no evidence for the existence of dark matter. The next measurable
length scale is the thickness of the Galactic disk, of order
300parsecs.  As noted above, the Kuijken/Gilmore analysis provides the
unexpected result that dark matter does not cluster on scale lengths
of a few hundred parsecs. Rotation curve analyses show however that
dark matter is gravitationally dominant on length scales of a few kpc
or so in all spiral, and perhaps elliptical, galaxies. What happens in
between?

The only intermediate scale length available for study is the 1kpc
scale size of the dwarf spheroidal galaxies, nine of which are
companion satellite galaxies of the Milky Way.

\subsection{Dark Matter in the Galactic Satellites}

The Galaxy's dwarf spheroidal satellites are very low surface
brightness, purely stellar, companion galaxies, each containing some
$10^6$ to $10^8$ stellar masses of stars, and with characteristic size
of order 1kpc, $10^{20}$m. The expected internal
kinematic velocity dispersion, assuming a purely stellar system, is
therefore $\sim 4-5$km/s. It was discovered in the early 1980's that
measured velocity dispersions are significantly higher, of order
10km/s, indicating dominance by very cold dark matter. A long
debate followed concerning the true precision of the kinematic data, and the
possible contribution to the apparent velocity dispersion from orbital
motions in unresolved stellar binaries. Recently, several independent
groups have confirmed the requisite quality of the kinematic data (eg
Hargreaves, Gilmore, Irwin \& Carter 1994), while extensive Monte
Carlo modeling has proven that unresolved binarism is not an important
factor (Hargreaves, Gilmore \& Annan 1996). The dwarf spheroidals are
in fact gravitationally dominated by dark matter throughout their
volume, with the dark matter density being several times the stellar
mass density even in the galactic core.

In 1994 the closest and the largest dwarf spheroidal, the Sagittarius
dwarf, was discovered only 15kpc from the centre of the Milky Way, by
Ibata, Gilmore \& Irwin (1994, 1995). This object is deep inside
the Galactic halo potential, (it is only twice as far from the
Galactic centre as is the Sun) and thereby undergoing strong tidal
stresses, and is also, by a substantial factor, the closest
dark-matter dominated system to the Sun. It provides an unrivalled
opportunity to test the true spatial distribution of dark matter, by
allowing calculation of its mass distribution independently from
stellar kinematics and from tidal survival arguments. 

This analysis has recently been completed by Ibata, Wyse, Gilmore,
Irwin \& Suntzeff (1997).  These authors conclude that the Sagittarius
dwarf galaxy is indeed dark matter dominated, and that its central
dark matter density is $\sim0.05~M_\odot$pc$^{-3} \equiv
1.5$GeV/cc. Interesting, and uniquely in this case, a length scale can
also be determined reliably, from the tidal radius of the system
imposed by the Galactic potential. This length scale is
$\sim1$kpc$\equiv 10^{20}$m. The mass density {\em profile} of the
dark matter can also be calculated, from tidal survival analysis. This
profile is flat, resembling a Heaviside function. Thus, one may
self-consistently, for the first time, associate a density and a
(tidally truncated) length scale to the dark matter which dominates
galaxies. A similar length scale has been suggested from (very
model-dependent) analyses of the rotation curves of dwarf spiral
galaxies.  Interestingly, this $10^{20}$m length scale is entirely consistent
with the absence of a substantial dark matter component associated
with the much smaller length scale of the Galactic disk, in excellent
agreement with the results summarised above.

\section {Conclusion}

In summary, the news is good for particle experimenters: all plausible
candidates for baryonic contributions to the Galactic dark matter can
be excluded, or at least very severely constrained. The only
alternatives to particle dark matter which have not yet been excluded require
astrophysically implausible scenarios. 

The dynamics of the Galaxy and its Satellites provide both a value for
the local dark matter density -- $\approx 
0.3$GeV/cc $\equiv 0.01M_\odot$pc$^{-3}$ -- and strong evidence that
it has an extended halo-like distribution, with smallest
characteristic length scale  $\sim 1kpc \equiv 10^{20}$m and maximum
observed mass density $\sim 0.05M_\odot$pc$^{-3} \equiv  1.5$GeV/cc.

\end{document}